\documentclass[preprint,aps,amsmath,byrevtex,prd,titlepage,
nofootinbib]{revtex4-1}

\usepackage{dcolumn}
\usepackage{amsmath}
\usepackage{amssymb}
\usepackage{bm}
\usepackage{hyperref}
\usepackage{graphicx}
\usepackage{slashed}

\newcommand{\beq}{\begin{equation}}
\newcommand{\eeq}{\end{equation}}
\newcommand{\eq}[1]{Eq.~(\ref{#1})}

\begin{document}

\title {Hard Three-Loop Corrections to Hyperfine Splitting in Positronium and Muonium}


\author {Michael I. Eides}
\altaffiliation[Also at ]{the Petersburg Nuclear Physics Institute,
Gatchina, St.Petersburg 188300, Russia}
\email[Email address: ]{eides@pa.uky.edu, eides@thd.pnpi.spb.ru}
\affiliation{Department of Physics and Astronomy,
University of Kentucky, Lexington, KY 40506, USA}
\author{Valery A. Shelyuto}
\email[Email address: ]{shelyuto@vniim.ru}
\affiliation{D. I.  Mendeleyev Institute for Metrology,
St.Petersburg 190005, Russia}

\begin{abstract}
We consider hard three-loop corrections to hyperfine splitting in muonium and positronium generated by the diagrams with closed electron loops. There are six gauge-invariant sets of such diagrams that generate corrections of order $m\alpha^7$. The contributions of these diagrams are calculated for an arbitrary electron-muon mass ratio without expansion in the small mass ratio. We obtain the formulae for contributions to hyperfine splitting that in the case of small mass ratio describe corrections for muonium  and  in the case of equal masses describe corrections for positronium. First few terms of the expansion of hard corrections in the small mass ratio were  earlier calculated for muonium analytically. We check numerically that the new results coincide with the sum of the known terms of the expansion in the case of small mass ratio. In the case of equal masses we obtain hard nonlogarithmic corrections  of order $m\alpha^7$ to hyperfine splitting in positronium.
\end{abstract}



\maketitle

\section{Introduction}

For many years hyperfine splitting (HFS) in muonium and positronium remains an active field of experimental and theoretical research. Results of highly accurate HFS measurements can be compared with the theoretical predictions of quantum electrodynamics obtained from the first principles without any adjustable parameters. Both experiment and theory have achieved very high accuracy. The experimental errors for HFS in muonium are now in the interval 16-51 Hz \cite{mbb,lbdd}, and a new measurement with the goal to reduce the error to about 10 Hz or to a few parts in $10^9$ is now planned at J-PARC \cite{sasaki,tanaka}. Current theoretical uncertainty of HFS in muonium is about 70-100 Hz, see, e.g., reviews in \cite{egs2001,egs2007,mtn2012}. Recent theoretical work on HFS in muonium concentrated on calculation of radiative-recoil corrections of order $\alpha^3(m/M)E_F$ that arise from the three-loop diagrams with closed electron and muon loops \cite{es2009prl,es2009pr,es2010jetp,es2013,es2014}. The goal of this work is to reduce the theoretical error below 10 Hz.

The hyperfine splitting in positronium is measured with the error bars at the level of 1-2 MHz \cite{mb73,rehw84,mills84,inaksyty2013}. There is a discrepancy about three standard deviations between the results of old and new experiments. New measurement of the positronium HFS splitting is now planned at J-PARC \cite{ishida2015}. All theoretical contributions to HFS in positronium of order $m\alpha^6$ and logarithmic corrections of order $m\alpha^7$ are already known, see, e.g., reviews in \cite{penin2014,bmp2014,af2014}. A new stage in the theory of positronium HFS was opened in \cite{bmp2014} where the one-photon annihilation contribution of order $m\alpha^7$ was calculated. This paper was soon followed by the works of Adkins and collaborators \cite{af2014,apswf2014}, who calculated contributions of the light-by-light scattering insertion in the scattering and annihilation channels.

Hard nonlogarithmic contributions to HFS in positronium of order $m\alpha^7$ are similar to the radiative and radiative-recoil corrections to HFS in muonium of orders $\alpha^2(Z\alpha)E_F$ and $\alpha^2(Z\alpha)(m/M)E_F$, respectively. We have calculated these corrections in muonium some time ago \cite{es2009prl,es2009pr,es2010jetp,es2013,es2014}.  The corrections in muonium are power series in the electron-muon mass ratio with the coefficients enhanced by large logarithms of this mass ratio. The goal of the old work on muonium was to calculate the  coefficients in this expansion, at least the factors before the logarithms, analytically. In the case of positronium the masses are equal and the hard corrections of order $m\alpha^7$ are pure numbers. We apply the approach developed for muonium to positronium. We consider an electromagnetically bound system of two particles with arbitrary masses $M$ and $m$, and obtain general  expressions for the hard corrections to HFS of order $m\alpha^7$ without expansion in the mass ratio of the constituents. We check numerically that in the case of a small mass ratio these formulae reproduce with high accuracy the sum of all already known terms in the expansion in the small mass ratio for muonium. We use the general expressions for the case of equal masses and calculate all hard three-loop contributions to HFS in positronium of order $m\alpha^7$ that are due to the diagrams with closed electron loops. The results of these calculations were reported in the rapid communication \cite{es2014r}. Below we present the details of the calculations in the general case of arbitrary mass ratio and in the special case of equal masses, for  positronium.

\section{Calculations}

\begin{figure}[htb]
\includegraphics
[height=1cm]
{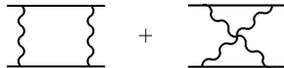}
\caption{\label{twoph}
Diagrams with two-photon exchanges}
\end{figure}

We start with the infrared divergent  contribution to HFS in muonium generated by the two-photon exchange diagrams in Fig.~\ref{twoph} calculated in the scattering approximation

\beq  \label{basicintini}
\begin{split}
\Delta E&=-\frac{Z\alpha}{\pi}E_F \frac{3mM}{16}\int \frac{d^4q}{i\pi^2q^4}L_{e,skel}^{\alpha\beta}(q)L_{\mu,skel,\alpha\beta}(-q)
\\
&=-\frac{Z\alpha}{\pi}E_F (2{mM})\int \frac{d^4q}{i\pi^2q^4}(2q^2+q_0^2)L^{(e)}_{skel}(q)L^{(\mu)}_{skel}(-q),
\end{split}
\eeq

\noindent
where

\beq \label{skelff}
L_{e,skel}^{\alpha\beta}(q)\equiv
-\frac{2q^2}{q^4-4m^2q^2_0}\gamma^{\mu}\hat q \gamma^{\nu}
=2L^{(e)}_{skel}\gamma^{\mu}\hat q \gamma^{\nu}
\eeq

\noindent
is the forward electron Compton scattering amplitude in the tree approximation (the skeleton electron-line factor), and $L_{\mu,skel}^{\alpha\beta}(q)$ is a similar amplitude for the muon. The Fermi energy is defined as $E_F=(8/3)(Z\alpha)^4m_r^3/(mM)$, where $m_r=mM/(m+M)$ is the reduced mass. In the case of equal masses, $M=m$, the Fermi energy $E_F$ turns into the leading nonannihilation contribution to HFS in positronium $E^{\rm Ps}_F=m\alpha^4/3$. The external electron and muon lines in the diagrams in Fig.~\ref{twoph} are on the mass shell and carry zero spatial momenta. In the second line in \eq{basicintini} we calculated projection of the matrix elements on HFS.

\noindent
After the Wick rotation and transition to four-dimensional spherical coordinates ($q_0=q\cos\theta$, $|\bm q|=q\sin\theta$) we obtain

\beq  \label{basicintmu}
\begin{split}
\Delta E&=
\frac{Z\alpha}{\pi}E_F
\frac{4mM}{\pi}\int_{0}^{\pi} {d\theta}\sin^2{\theta} \int_{0}^{\infty}
{dq^2}(2+\cos^2{\theta})L^{(e)}_{skel}L^{(\mu)}_{skel}
\\
&=\frac{Z\alpha}{\pi}E_F
\frac{4mM}{\pi}\int_{0}^{\pi} {d\theta}\sin^2{\theta} \int_{0}^{\infty}
{dq^2}\frac{2+\cos^2{\theta}}{(q^2+4m^2\cos^2{\theta})(q^2+4M^2\cos^2{\theta})}
\\
&\equiv
\frac{Z\alpha}{\pi}E_F \frac{mM}{M^2-m^2}\int_0^\infty dq^2f_\mu(q),
\end{split}
\eeq

\noindent
where at the last step we rescaled the integration momentum $q\to qm$. The dimensionless weight function  $f_\mu(q)$ in terms of an auxiliary function

\beq
f(q)=-\frac{1}{4}+\frac{\sqrt{q^2+4}}{4q}-\frac{2\sqrt{q^2+4}}{q^3}
\eeq

\noindent
has the form

\beq
f_\mu(q)=f(q)-4\mu^2 f(2\mu q),
\eeq

\noindent
where $\mu=m/(2M)$.

In the case of positronium $M\to m$ and the weight function simplifies

\beq
{\frac{mM}{M^2-m^2}f_\mu(q)}_{|M\to m}\to \frac{16+2q^2+q^4-q^3\sqrt{q^2+4}}{4q^3\sqrt{q^2+4}}\equiv f_p(q).
\eeq

\noindent
Respectively, the skeleton integral in \eq{basicintmu} in the case of positronium turns into

\beq  \label{basicint}
\begin{split}
\Delta E&=
\frac{\alpha}{\pi}E^{\rm Ps}_F
\frac{4m^2}{\pi}\int_{0}^{\pi} {d\theta}\sin^2{\theta} \int_{0}^{\infty}
{dq^2}L_{e,skel}^2(2+\cos^2{\theta})
\\
&=\frac{\alpha}{\pi}E^{\rm Ps}_F
\frac{4m^2}{\pi}\int_{0}^{\pi} {d\theta}\sin^2{\theta} \int_{0}^{\infty}
{dq^2}\frac{2+\cos^2{\theta}}{(q^2+4m^2\cos^2{\theta})^2}
\equiv
\frac{\alpha}{\pi}E^{\rm Ps}_F \int_0^\infty dq^2f_p(q).
\end{split}
\eeq

The integrals in \eq{basicintmu} and \eq{basicint} are  sums of an infrared linearly divergent integral and a finite one. In a more accurate approximation (with the off mass shell external fermion lines) the linear divergence is cutoff at the characteristic atomic scale $\sim m\alpha$ and turns into a contribution of a lower order in $\alpha$. The remaining finite part of the integral originates at hard integration momenta $\sim m$ (or in the interval from $m$ to $M$ in the case of unequal masses) and generates a contribution of order $\alpha E_F$. In the case of unequal masses, for muonium, the linearly infrared divergent contribution turns into the leading nonrecoil Fermi contribution $E_F$ to HFS, while the finite part generates the leading recoil correction of order $\alpha(m/M)E_F$, see, e.g., \cite{egs2001,egs2007}. Let us emphasize that due to the linear (as opposed to logarithmic) nature of the apparent divergence it leaves no finite remnant of order $\alpha E_F$ and should be simply thrown away. No need in matching of high and low integration momenta arises.

Six gauge-invariant sets of diagrams in Figs.~\ref{twooneloop} - \ref{onelopradph} and in Figs.~\ref{polinrad} - \ref{combo} generate hard radiative corrections of order $m\alpha^7$ that are due to the graphs with closed electron loops\footnote{All gauge-invariant sets of diagrams include the graphs with the crossed exchanged photons that we do not show explicitly.} . All these diagrams can be interpreted as the results of radiative insertions in the skeleton diagrams with two-photon exchanges in Fig.~\ref{twoph}. It is well known that insertion of radiative corrections suppresses the low integration momentum region, see, e.g., \cite{egs2001,egs2007,egs1991}. Hence, all diagrams in Figs.~\ref{twooneloop} - \ref{onelopradph} and in Figs.~\ref{polinrad} - \ref{combo} are infrared convergent\footnote{Linearly infrared divergent contributions due to the anomalous magnetic moment should be subtracted from radiative corrections in Figs.~\ref{polinrad} and Fig.~\ref{combo}, see more on this below.}. Moreover, the characteristic integration momenta in these diagrams are hard (of order $\sim m$ or in the interval from $m$ to $M$ in the case of unequal masses) and are much larger than the atomic momenta of order $\sim m\alpha$, what justifies validity of the scattering approximation for their calculation. This is exactly the approximation we used above in calculation of the contribution of the skeleton diagrams in Fig.~\ref{twoph}, and all corrections calculated below are obtained by some modifications of the basic integrals in \eq{basicintmu} and \eq{basicint}.

\subsection{Analytic Results for One- and Two-Loop Polarization Insertions in the Exchanged Photons}

\begin{figure}[htb]
\includegraphics
[height=1.5cm]
{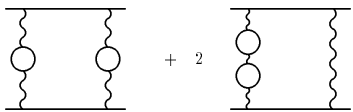}
\caption{\label{twooneloop}
Diagrams with two one-loop polarization insertions}
\end{figure}

Consider first the diagrams in Fig.~\ref{twooneloop}  with two one-loop polarization loops. Insertion of a polarization operator in a photon line with momentum  $q$ (all momenta below are measured in units of the electron mass) reduces to the replacement in the photon propagator

\beq \label{replonelppol}
\frac{1}{q^2}\to \frac{\alpha}{\pi}I_1(q),
\eeq

\noindent
where $(\alpha/\pi)I_1(q)$ is  the well known representation of the one-loop vacuum polarization \cite{schw1973}

\beq \label{onelppol}
\frac{\alpha}{\pi}I_1(q)=\frac{\alpha}{\pi}
\int_0^1dv\frac{v^2(1-\frac{v^2}{3})}{1-v^2}\frac{1}{q^2+\frac{4}{1-v^2}}.
\eeq

\noindent
We see that a photon line that carries a polarization loop has a natural interpretation as a propagator of a massive photon with mass squared $\lambda^2=4/(1-v^2)$. According to \eq{onelppol} this propagator should be integrated over $v$ with the weight $(\alpha/\pi)v^2(1-v^2/3)/(1-v^2)$.


The contribution of the diagrams in Fig.~\ref{twooneloop} is obtained by  insertion of the one-loop photon polarization squared $(\alpha/\pi)^2q^4I_1^2(q)$ in the integrands  in \eq{basicintmu} and \eq{basicint}. Due to nonsingular behavior of the polarization operator at $q^2\to0$ we obtain convergent integrals where the effective integration momenta are hard, of order $\sim m$ (or in the interval from $\sim m$ to $M$ in the case of unequal masses). Then in the general case of unequal masses the contribution to HFS has the form

\beq \label{onepoloppmu}
\Delta E=3\frac{\alpha^2(Z\alpha)}{\pi^3}E_F \frac{mM}{M^2-m^2}\int_0^\infty dq^2f_\mu(q)q^4I_1^2(q),
\eeq

\noindent
where the factor 3 before the integral has the combinatorial origin.
We checked numerically that in the small mass ratio limit this integral
reproduces the sum of all known analytically terms \cite{eks1989,egs2001pr} of the expansion of this contribution in the small mass ratio.

In the case of positronium the integral in \eq{onepoloppmu} reduces to (compare \eq{basicint})

\beq
\Delta E_1=3\frac{\alpha^3}{\pi^3}E^{\rm Ps}_F \int_0^\infty dq^2f_p(q)q^4I_1^2(q),
\eeq

\noindent
and after computation we obtain the contribution to HFS of the diagrams with two one-loop polarization insertions in Fig.~\ref{twooneloop}

\beq \label{onepol}
\Delta E_1=\left(\frac{6\pi^2}{35} -
\frac{8}{9}\right)\frac{\alpha^3}{\pi^3}E^{\rm Ps}_F
=0.803~043~294\frac{\alpha^3}{\pi^3}E^{\rm Ps}_F.
\eeq

\begin{figure}[htb]
\includegraphics
[height=1.5cm]
{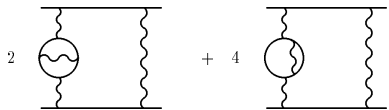}
\caption{\label{onetwoloop}
Diagrams with two-loop polarization insertions}
\end{figure}

The contribution of the two-loop vacuum polarization in Fig.~\ref{onetwoloop} can be obtained by the insertion of the two-loop photon polarization $(\alpha^2/\pi^2)q^2I_2(q)$ \cite{ks1955,schw1973} in the integrands in \eq{basicintmu} and \eq{basicint}

\beq \label{kssch}
\begin{split}
\left(\frac{\alpha}{\pi}\right)^2I_2(q)
&
=\frac{2}{3}
\left(\frac{\alpha}{\pi}\right)^2\int_0^1dv\frac{v}{4+q^2(1-v^2)}
\Biggl\{(3-v^2)(1+v^2)
\Biggl[{\rm Li}_2\left(-\frac{1-v}{1+v}\right)
\\
&
+2{\rm Li}_2\left(\frac{1-v}{1+v}\right)
+\frac{3}{2}\ln\frac{1+v}{1-v}\ln\frac{1+v}{2}
-\ln\frac{1+v}{1-v}\ln v\Biggr]
\\
&
+\left[\frac{11}{16}(3-v^2)(1+v^2)+\frac{v^4}{4}\right]\ln\frac{1+v}{1-v}
\\
&
+\left[\frac{3}{2}v(3-v^2)\ln\frac{1-v^2}{4}-2v(3-v^2)\ln v\right]
+\frac{3}{8}v(5-3v^2)\Biggr\},
\end{split}
\eeq

\noindent
where the dilogarithm is defined as ${\rm Li}_2(z)=-\int_0^1dt{\ln(1-zt)}/{t}$.

In the case of unequal masses the integral for the contribution to HFS of the diagrams   with the two-loop polarization in Fig.~\ref{onetwoloop} loop  has the form

\beq \label{kstwomu}
\Delta E=2\frac{\alpha^2(Z\alpha)}{\pi^3}E_F\frac{mM}{M^2-m^2} \int_0^\infty dq^2f_\mu(q)q^2I_2(q),
\eeq

\noindent
where the factor 2 before the integral is due to combinatorics. Again, due to nonsingular behavior of the two-loop polarization at small $q^2\to0$  the integral in \eq{kstwomu} is convergent, and typical integration momenta are hard, in the interval from $m$ to $M$. We checked numerically that in the small mass ratio case the integral in \eq{kstwomu} coincides with the sum of the known terms  \cite{eks1989,egs2001pr} of the expansion of this contribution to HFS in the small mass ratio.

In the case of equal masses, for positronium, the contribution  to HFS of the diagrams in Fig.~\ref{onetwoloop}   reduces to the integral

\beq \label{kstwo}
\Delta E_2=2\frac{\alpha^3}{\pi^3}E^{\rm Ps}_F \int_0^\infty dq^2f_p(q)q^2I_2(q).
\eeq

\noindent
This integral admits an analytic calculation, and we obtain

\beq \label{twpol}
\Delta E_2=\left[-\frac{217}{30}\zeta{(3)} + \frac{28\pi^2}{15}\ln{2}+
\frac{\pi^2}{675} +  \frac{403}{360}\right]\frac{\alpha^3}{\pi^3}E^{\rm Ps}_F
=5.209~219~614\frac{\alpha^3}{\pi^3}E^{\rm Ps}_F.
\eeq

\subsection{One-Loop Electron Factor and One-Loop Polarization Insertion in the Exchanged Photon}

\begin{figure}[htb]
\includegraphics
[height=2cm]
{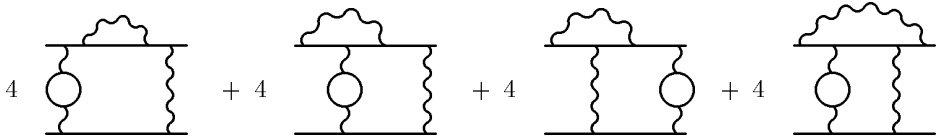}
\caption{\label{onelopradph}
Diagrams with one-loop polarization and radiative photon insertions}
\end{figure}

\begin{figure}[htb]
\includegraphics
[height=1cm]
{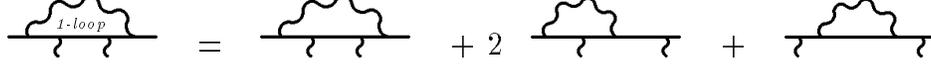}
\caption{\label{ff}
One-loop fermion factor}
\end{figure}

The diagrams in Fig.~\ref{onelopradph} are obtained from the skeleton diagrams in Fig.~\ref{twoph} by one-loop radiative insertions in one of the exchanged photons and one of the fermion lines\footnote{Multiplicity factors in these diagrams correspond to the case of positronium, not muonium.}. To describe these radiative insertions it is convenient to introduce the one-loop electron factor that is defined as a gauge invariant sum of the diagrams in Fig.~\ref{ff} where the external electron lines are on-shell  and carry zero spatial momenta (plus the diagrams with the exchanged external photon vertices). Physically the electron factor is a sum of one-loop corrections to the spin-dependent amplitude of the virtual forward Compton scattering.

The gauge invariant electron factor $\widetilde L_{\mu\nu}$ can be written as a sum of two gauge invariant terms $\widetilde L_{\mu\nu}=L_{\mu\nu}+L_{\mu\nu}^{(a)}$, where the term  $L_{\mu\nu}^{(a)}$ is the contribution of the anomalous magnetic moment (for more details see, e.g., \cite{egs2004,es2014_d}). The multiloop electron factors also can be written as sums of two gauge invariant terms.  Representation of the electron factor in the form of a sum of two gauge invariant terms is convenient for calculations because these terms have different behavior at low virtual photon momenta. According to the generalized low-energy theorem (see, e.g., \cite{egs2001,egs2007}) all terms linear in the small photon momentum $q$ are due to the term $L_{\mu\nu}^{(a)}$, while the term $L_{\mu\nu}$ decreases at least as $q^2$ at small $q^2$. This different low-energy behavior determines the structure of the integrals for the contributions to hyperfine splitting. In the case of the diagrams in Fig.~\ref{polinrad} and in Fig.~\ref{combo} the contributions to HFS generated by the term $L_{\mu\nu}^{(a)}$ are of lower order in $\alpha$ than the apparent order of a diagram. Technically presence of the previous order contribution reveals itself as a linear infrared divergence of an integral calculated in the scattering approximation.

In the diagrams in Fig.~\ref{onelopradph} the skeleton fermion line in Fig.~\ref{twoph} is effectively replaced by the one-loop fermion factor $\widetilde L_{\mu\nu}$ in Fig.~\ref{ff}, what can be described by  the substitution

\beq \label{elfact}
L_{e,skel}^{\mu\nu}(q)
\to \widetilde L^{\mu\nu}(q)=
2\frac{\alpha}{4\pi}\left\{ \gamma^{\mu}\hat q \gamma^{\nu}
 {\widetilde L}_{\mbox{\tiny I}}(q^2, q^2_0) +q_0
\left[\gamma^{\mu}\gamma^{\nu} -\frac{q^{\mu}\hat q
\gamma^{\nu}+\gamma^{\mu}\hat q q^{\nu} }{q^2}\right]  {\widetilde
L}_{\mbox{\tiny II}}(q^2, q^2_0) \right\}.
\eeq

\noindent
where ${\widetilde L}_{\mbox{\tiny I(II)}}$ are scalar form factors. The scalar form factors  ${\widetilde L}_{\mbox{\tiny I(II)}}$ have the form

\beq \label{totintan}
{\widetilde
L}_{\mbox{\tiny I}}=L_{\mbox{\tiny I}}+L_A,\qquad
{\widetilde
L}_{\mbox{\tiny II}}=L_{\mbox{\tiny II}}-L_A,
\eeq

\noindent
where the  scalar form factors ${L}_{\mbox{\tiny I(II)}}$ and $L_A$ correspond to $L_{\mu\nu}$ and $L_{\mu\nu}^{(a)}$, respectively. The factor 2 before the braces arises because we normalize  the scalar form factors like the skeleton one in \eq{skelff}, and the factor $\alpha/(4\pi)$ is due to the one-loop integration in the fermion factor.

The one-loop electron factor $L_{\mu\nu}$ with the subtracted contribution of the anomalous magnetic moment enters calculations of the two-loop radiative-recoil corrections to HFS in muonium, and we had derived an explicit integral representations for the respective scalar form factors ${L}_{\mbox{\tiny I(II)}}$ long time ago \cite{beks1989,egs1998,egs2003}. After the Wick rotation, rescaling of the integration momentum $q\to qm$, and transition to the four-dimensional spherical coordinates the form factors can be written as

\beq
\begin{split}
L_{\mbox{\tiny I}}(q^2,\cos^2\theta)
&=\int_0^1 {dx} \int_0^x
{dy}\Biggl\{
\frac{(q^2+a^2)\left[(q^2+a^2)^2-12b^2q^2\cos^2\theta\right]}
{\left[(q^2+a^2)^2+4b^2q^2\cos^2\theta\right]^3}
\left(c_1 q^2\sin^2\theta + c_2  q^4\right)
\\
&-\frac{(q^2+a^2)^2-4b^2q^2\cos^2\theta}{\left[(q^2+a^2)^2
+4b^2q^2\cos^2\theta\right]^2}c_3 q^2 + \frac{(q^2+a^2)4b
q^2\cos^2\theta}{\left[(q^2+a^2)^2+4b^2q^2\cos^2\theta\right]^2}2c_4\Biggr\},
\\
L_{\mbox{\tiny II}}(q^2,\cos^2\theta)
&=\int_0^1 {dx} \int_0^x
{dy}\Biggl\{ \frac{(q^2+a^2)
4b}{\left[(q^2+a^2)^2+4b^2q^2\cos^2\theta\right]^2}  c_5 q^2
\\
&-
\frac{(q^2+a^2)^2-4b^2q^2\cos^2\theta}{\left[(q^2+a^2)^2+4b^2q^2\cos^2\theta\right]^2}
2 c_6 q^2 + \frac{2b}{(q^2+a^2)^2+4b^2q^2\cos^2\theta} c_7{q^2}
\Biggr\},
\end{split}
\eeq

\noindent
where $a^2 = {x^2}/{y(1 - y)}$, $b = ({1 - x})/({1 - y})$, and the coefficient functions $c_i$ are collected in Table \ref{table1}.

\begin{table}
\caption{\label{table1}Coefficients in the Electron-Line Factor}
\begin{ruledtabular}
\begin{tabular}{ll}
$\mbox{c}_{1}$& $\frac{16}{y(1-y)^3} \left[(1-x)(x-3y)-2y\ln{x}\right]$
\\
$\mbox{c}_{2}$      &   $\frac{4}{y(1-y)^3} \left[-(1-x)\left(x - y - \frac{2y^2}{x}\right)+2\left(x - 4y + \frac{4y^2}{x}\right)  \ln{x}\right]$
\\
$\mbox{c}_{3}$     &   $\frac{1}{y(1-y)^2} \left[1 - 6x - 2x^2
-\frac{y}{x}\left(26 - \frac{6y}{x} - 37x -2x^2 + 12xy + 16 \ln{x}\right)\right]$
\\
$\mbox{c}_{4}$   & $\frac{1}{y(1-y)^2} \left(2x - 4x^2 - 5y + 7xy \right)$
\\
$\mbox{c}_{5}$   & $\frac{1}{y(1-y)^2} \left( 6x - 3x^2 - 8y + 2xy \right)$
\\
$\mbox{c}_{6}$   & $-b^2\frac{x - y}{x^2}$
\\
$\mbox{c}_{7}$  & $2  \frac{1-x}{x}$
\end{tabular}
\end{ruledtabular}
\end{table}

The scalar form factor $L_A$ is proportional to the respective skeleton form factor $L_{skel}$ in \eq{skelff}. After the Wick rotation and in terms of the dimensionless integration momentum $q$ it has the form

\beq \label{anformf}
L_{\mbox{\tiny A}}=\frac{2}{q^2+4\cos^2\theta}=2L_{skel}.
\eeq

In the case of unequal masses an analytic expression for the contribution to HFS of the diagrams in Fig.~\ref{onelopradph}  is obtained by modification of the skeleton integral in \eq{basicintmu}. First, we replace the skeleton factor in the integrand

\beq
(2+\cos^2\theta)L^{(e)}_{skel}\to\frac{\alpha}{4\pi}\left[(2+\cos^2{\theta}){\widetilde
L}_{\mbox{\tiny I}}-3\cos^2\theta {\widetilde
L}_{\mbox{\tiny II}}\right].
\eeq

\noindent
The factor $\alpha/(4\pi)$ comes from the substitution in \eq{elfact}, and the term with ${\widetilde L}_{\mbox{\tiny II}}$ arises because the one-loop electron factor in \eq{elfact} contains an additional spinor structure in comparison with the skeleton one in \eq{skelff}.

Second, we need to account for the polarization loops in Fig.~\ref{onelopradph} and insert the term $2q^2I_1(q)$ in the integrand in \eq{basicint}. The factor 2 is due to two ways to insert the polarization operator in one of the exchanged photons. The polarization operator $q^2I_1(q)$ decreases like $q^2$ at small $q$, and we obtain an infrared convergent integral with hard characteristic integration momenta of order $m$ (or in the interval from $m$ to $M$ in the case of unequal masses). Due to suppression of the small integration momenta the anomalous magnetic moment  in the diagrams in Fig.~\ref{onelopradph}  gives contribution on par with the other terms in the one-loop electron factor $\widetilde L^{\mu\nu}(q)$.

Then the contribution to HFS of the diagrams in Fig.~\ref{onelopradph}  in the case of unequal masses has the form

\beq \label{polelfmu}
\Delta E=
\frac{\alpha^2(Z\alpha)}{\pi^3}E_F\frac{M}{m}\frac{2}{\pi}\int_0^\infty dq^2q^2I_1(q)\int_0^\pi d\theta\sin^2\theta
L^{(\mu)}_{skel}\left[(2+\cos^2\theta){\widetilde
L}_{\mbox{\tiny I}} -3\cos^2\theta{\widetilde L}_{\mbox{\tiny
II}}\right].
\eeq

The leading terms of the expansion of the contribution to HFS in \eq{polelfmu} in the small mass ratio are already known for some time \cite{eks1989,egs2003,egs2005can}

\beq
\begin{split}
\Delta E=
\frac{\alpha^2(Z\alpha)}{\pi^3}E_F
&\Biggl[
\pi^2
\left(
-\frac{4}{3}\ln^2\frac{1+\sqrt{5}}{2}-\frac{20}{9}
\sqrt{5}\ln\frac{1+\sqrt{5}}{2}-\frac{64}{45}\ln2
+\frac{\pi^2}{9}+\frac{3}{8}+\frac{1043}{675}
\right)
\\
&
+\frac{m}{M}\left(\frac{5}{2}\ln^2\frac{M}{m}+\frac{10}{3}\ln\frac{M}{m}
+11.41788\right)\biggr].
\end{split}
\eeq

\noindent
We have checked numerically that the integral in \eq{polelfmu} coincides with this analytical result in the case of small mass ratio.

In the case of equal masses there is an extra factor 2 before the diagrams in Fig.~\ref{onelopradph}. This factor arises because now there are two ways to insert the fermion factor in one of the lepton lines. Hence, the respective contribution to HFS is described by the doubled integral in \eq{polelfmu} at $M=m$. Then we obtain the contribution of the diagrams in Fig.~\ref{onelopradph} to HFS in positronium in the form

\beq \label{polelf}
\begin{split}
\Delta E_3&=
\frac{\alpha^3}{\pi^3}E_F^{Ps} \frac{4}{\pi}\int_0^\infty dq^2q^2I_1(q)\int_0^\pi d\theta\sin^2\theta
L_{skel}\left[(2+\cos^2\theta){\widetilde
L}_{\mbox{\tiny I}} -3\cos^2\theta{\widetilde L}_{\mbox{\tiny
II}}\right]
\\
&=-1.287~09~(1)~\frac{\alpha^3}{\pi^3}E_F^{Ps}.
\end{split}
\eeq

\subsection{One-Loop Polarization Insertion in the Electron Factor}

\begin{figure}[htb]
\includegraphics
[height=2cm]
{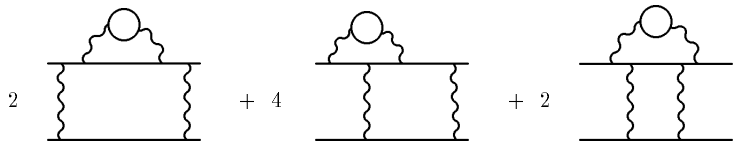}
\caption{\label{polinrad}
Diagrams with one-loop polarization insertions in radiative photons}
\end{figure}

Consider now the diagrams in Fig.~\ref{polinrad} with the one-loop polarization insertions in the radiative photon\footnote{Multiplicity factors in these diagrams correspond to the case of positronium, not muonium.}. Effectively these diagrams contain a massive radiative photon, see \eq{replonelppol} and \eq{onelppol}. In principle, the respective electron factor can be obtained from the one-loop electron factor in \eq{elfact} by restoring the radiative photon mass squared $\lambda^2=4m^2/(1-v^2)$, followed by integration over $v$ with the weight $(\alpha/\pi)v^2(1-v^2/3)/(1-v^2)$. However, the relatively compact expression in \eq{elfact} is a result of numerous cancellations in the integrand between the contributions from different diagrams in Fig.~\ref{ff}, and technically it is much easier to start calculation of the two-loop electron factor in Fig.~\ref{polinrad} from scratch. We consider this electron factor as a sum of the contributions corresponding to the separate diagrams in  Fig.~\ref{polinrad} with the self-energy, vertex and spanning photon insertions in the electron line. Each of these terms is calculated as a one-loop diagrams with a massive photon and then integrated over the auxiliary parameter $v$ as we just explained.

The only subtlety in further calculations is connected with the diagrams with the vertex correction in Fig.~\ref{polinrad}.  All entries in the two-loop fermion factor except the two-loop anomalous magnetic moment carry at least one extra power of $q^2$ at $q^2\to0$ in comparison with the skeleton electron factor. One can separate the contribution to the two-loop anomalous magnetic moment from the two-loop vertex in the second diagram  in Fig.~\ref{polinrad} in a gauge invariant way, like we separated the one-loop anomalous magnetic moment from the one-loop electron factor $L_{\mu\nu}$ above. The two-loop  anomalous magnetic moment term in the second diagram in Fig.~\ref{polinrad} generates a linearly infrared divergent contribution to HFS. This linear infrared divergence that is cutoff at the characteristic atomic scale $\sim m\alpha$ indicates that the anomalous magnetic moment generates a contribution to HFS of the previous order in $\alpha$. This correction of order $m\alpha^6$ is already accounted for in earlier calculations and we should simply delete the apparently divergent term that generates it. To get rid of the spurious divergence we subtract the gauge invariant term with the two-loop anomalous magnetic moment from the two-loop electron vertex in Fig.~\ref{polinrad}. The subtracted two-loop electron factor in Fig.~\ref{polinrad} can be written in terms of scalar two-loop  form factors $L^{(2)}_{\mbox{\tiny I,II}}$

\beq \label{twloppff}
L^{(2)}_{\mbox{\tiny I,II}}=L^{(2,\Sigma)}_{\mbox{\tiny I,II}}+2L^{(2,\Lambda)}_{\mbox{\tiny I,II}}+L^{(2,\Xi)}_{\mbox{\tiny I,II}}
\eeq

\noindent
similar to the one-loop form factors $L_{\mbox{\tiny I,II}}$ in \eq{totintan}.
Unlike the one-loop form factors ${\widetilde
L}_{\mbox{\tiny I,II}}$ in \eq{elfact} these two-loop form factors do not include contributions of the anomalous magnetic moment.

We have derived explicit expressions for the two-loop form factor in \eq{twloppff} in \cite{es2010jetp,eks1990}, where we calculated nonrecoil and radiative-recoil correction to HFS in muonium due to the diagrams in Fig.~\ref{polinrad}. These expressions are rather cumbersome and we will not reproduce them here. The contribution to HFS of the diagrams in Fig.~\ref{polinrad} in the case of unequal masses has the form

\beq \label{polinradmu}
\Delta E=
\frac{\alpha^2(Z\alpha)}{\pi^3}E_F\frac{M}{m}\frac{1}{\pi}\int_0^\infty dq^2\int_0^\pi d\theta\sin^2\theta
L^{(\mu)}_{skel}\left[(2+\cos^2\theta)L^{(2)}_{\mbox{\tiny I}} -3\cos^2\theta L^{(2)}_{\mbox{\tiny II}}\right].
\eeq

\noindent
A few leading terms of the expansion of this contribution to HFS in the small mass ratio were calculated earlier

\beq
\Delta E=\frac{\alpha^2(Z\alpha)}{\pi^3}E_F\left\{-0.310~742~\pi^2
+\frac{m}{M}\left[\frac{3}{4}\ln^2\frac{M}{m}+\left(\pi^2-\frac{53}{6}\right)\ln\frac{M}{m}
+7.08072\right]\right\}.
\eeq

\noindent
We have checked numerically that the expression in \eq{polinradmu} derived for arbitrary masses reproduces the expansion above in the case of small mass ratio.

In the case of equal masses the contribution of the diagrams in Fig.~\ref{polinrad} to HFS  reduces to

\beq
\Delta E_4=\frac{\alpha^3}{\pi^3}E_F^{Ps} \frac{2}{\pi}\int_0^\infty dq^2\int_0^\pi d\theta\sin^2\theta
L^{(\mu)}_{skel}\left[(2+\cos^2\theta)L^{(2)}_{\mbox{\tiny I}} -3\cos^2\theta L^{(2)}_{\mbox{\tiny II}}\right],
\eeq

\noindent
where an extra factor 2 before the integral (in comparison with \eq{polinradmu}) arises because we can insert the two-loop electron factor in either of the fermion lines.

After calculations we obtain

\beq \label{polinrd}
\Delta E_4=-3.154~41~(1)~\frac{\alpha^3}{\pi^3}E_F^{Ps}.
\eeq

\subsection{Light-by-Light Scattering Contribution}

\begin{figure}[htb]
\includegraphics
[height=2cm]
{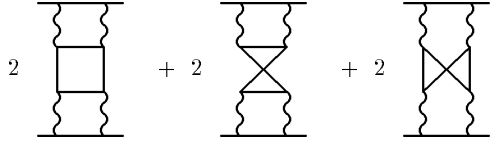}
\caption{\label{lbl}
Diagrams with light-by-light scattering insertions}
\end{figure}

Due to gauge invariance the light-by-light scattering block fast decreases with the momenta of the external (virtual) photons. Therefore, effectively all integration momenta in the diagrams in Fig.~\ref{lbl} are hard, of order of the electron mass (or in the interval from $m$ to $M$ in the case of unequal masses).

The contribution of the light-by-light scattering block to HFS in the general case of unequal masses has the form \cite{es2014,es2014_d}

\beq \label{lblnanmu}
\Delta E=
\frac{\alpha^2(Z\alpha)}{\pi^3} E_F\frac{3M^2}{32\pi}\int_0^\infty \frac{dq^2}{q^2}\int_0^\pi d\theta\sin^2\theta\frac{T(q^2,\cos^2\theta)}
{m^2q^2+4M^2\cos^2\theta}.
\eeq

\noindent
The explicit integral representation for the function  $T(q^2,\cos^2\theta)$ can be found in \cite{es2014}.

The first terms of the expansion of this contribution to HFS in the small mass ratio were calculated during the years \cite{eks1991,eks1989log,es2013,es2014}

\beq \label{expanlblmu}
\begin{split}
\Delta E=
\frac{\alpha^2(Z\alpha)}{\pi^3}E_F
&\Biggl\{-0.472~514~(1)~\pi^2
\\
&
+\frac{m}{M}\left[\frac{9}{4}\ln^2{\frac{{M}}{m}}
+ \biggl(-3\zeta{(3)}- \frac{2\pi^2}{3} + \frac{91}{8}\biggr) \ln{\frac{{M}}{m}} +5.9949(1)\right]\Biggr\}.
\end{split}
\eeq

\noindent
We have checked numerically that the general expression in \eq{lblnanmu} coincides with the sum in \eq{expanlblmu} in the case of small mass ratio.

In the case of equal masses the integral in \eq{lblnanmu} reduces to\footnote{There is a misprint in the respective expression in Eq.(14) in \cite{es2014r}.} 

\beq
\Delta E_5=
\frac{\alpha^3}{\pi^3} E^{Ps}_F\frac{3}{32\pi}\int_0^\infty \frac{dq^2}{q^2}\int_0^\pi d\theta\sin^2\theta\frac{T(q^2,\cos^2\theta)}
{q^2+4\cos^2\theta}.
\eeq

\noindent
Calculation of this contribution to HFS in positronium  proceeds exactly like calculation of the respective nonlogarithmic radiative-recoil correction to HFS in muonium in \cite{es2014} and we obtain

\beq \label{lbln}
\Delta E_5=-0.706~27~(5)~\frac{\alpha^3}{\pi^3}E^{Ps}_F,
\eeq

\noindent
what coincides with the result first obtained in \cite{af2014}.

\subsection{Two-One Loop Electron Factors}

\begin{figure}[htb]
\includegraphics
[height=2cm]
{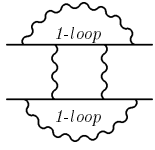}
\caption{\label{combo}
Diagrams with one-loop radiative photon insertions in both fermion lines}
\end{figure}

The diagrams in Fig.~\ref{combo} contain the one-loop fermion factors from \eq{elfact} in both fermion lines. Naively, the contribution of these diagrams to HFS can be obtained from the skeleton integral in \eq{basicintmu} by the replacement

\beq \label{naivffpr}
\begin{split}
L^{(e)}_{skel}L^{(\mu)}_{skel}(2+\cos^2{\theta})\to\left(\frac{\alpha}{4\pi}\right)^2
&\biggl[(2+\cos^2{\theta}){\widetilde
L}^{(e)}_{\mbox{\tiny I}}{\widetilde
L}^{(\mu)}_{\mbox{\tiny I}}-3\cos^2\theta \left({\widetilde
L}^{(e)}_{\mbox{\tiny I}}{\widetilde
L}^{(\mu)}_{\mbox{\tiny II}}+{\widetilde
L}^{(e)}_{\mbox{\tiny II}}{\widetilde
L}^{(\mu)}_{\mbox{\tiny I}}\right)
\\
&
+\cos^2\theta(1+2\cos^2{\theta}){\widetilde
L}^{(e)}_{\mbox{\tiny II}}{\widetilde
L}^{(\mu)}_{\mbox{\tiny II}}
\biggr],
\end{split}
\eeq

\noindent
where the terms on the right hand side arise after calculation of the projection of the product of two electron factors (see \eq{elfact}) on the HFS structure.

The scalar form factors ${\widetilde L}^{(e,\mu)}_{\mbox{\tiny I,II}}$  include terms with the scalar form factors $L^{(e,\mu)}_A$ arising due anomalous magnetic moments, see \eq{totintan}. Therefore, each product of the scalar functions in the square brackets on the right hand side of \eq{naivffpr} contains the term $L^{(e)}_AL^{(\mu)}_A$. As we already mentioned the form factors $L^{(e,\mu)}_{\mbox{\tiny I,II}}$ decrease at least as $q^2$ at $q^2\to0$ relative to the skeleton form factors, while the form factors $L^{(e,\mu)}_A$ behave exactly like the skeleton form factors, see  \eq{anformf}. Hence, each integral of $L^{(e)}_AL^{(\mu)}_A$ is a sum of a linearly infrared divergent and finite contributions, compare with the skeleton integral in \eq{basicintmu}. In a more accurate calculation the linearly infrared divergent contribution would be cutoff at the atomic scale $\sim m\alpha$  and would generate a correction of lower order in $\alpha$. It should be simply subtracted, while we need to preserve the finite part of the integral that generates correction of order $m\alpha^7$. In the general case of different masses (for example, for muonium) the finite part is a recoil contribution and it was calculated in \cite{egs2004}. It is equal to $(9/16)(mM)/(M^2-m^2)\ln(M^2/m^2)$ up to a normalization factor. Hence, subtraction of the linearly infrared divergent contribution due to terms with $L^{(e)}_AL^{(\mu)}_A$ in \eq{naivffpr} reduces to addition of $(9/16)(mM)/(M^2-m^2)\ln(M^2/m^2)$ to the respective contribution to HFS instead of all terms on the right hand side in \eq{naivffpr} that proportional to $L^{(e)}_AL^{(\mu)}_A$. After this replacement of the infrared divergent part we obtain a convergent integral with hard characteristic integration momenta in the interval from $m$ to $M$. The contribution to HFS of the diagrams in Fig.~\ref{combo} in the case of unequal masses has the form

\beq \label{combogen}
\begin{split}
\Delta E
&
=\frac{\alpha(Z^2\alpha)(Z\alpha)}{\pi^3}E_F\Biggl\{\frac{M}{m} \frac{1}{4\pi} \int_0^\infty {dq^2\int_0^\pi d\theta\sin^2\theta d\theta}
\biggl[(2+\cos^2\theta )
\\
&
\times
\left(L^{(e)}_{\mbox{\tiny I}}L^{(\mu)}_{\mbox{\tiny I}}+L^{(e)}_AL^{(\mu)}_{\mbox{\tiny I}}+L^{(\mu)}_AL^{(e)}_{\mbox{\tiny I}}\right)
\\
&
-3\cos^2\theta\left(L^{(e)}_{\mbox{\tiny I}}L^{(\mu)}_{\mbox{\tiny II}}+L^{(e)}_A L^{(\mu)}_{\mbox{\tiny II}}
-L^{(e)}_{\mbox{\tiny I}}L^{(\mu)}_A
+L^{(e)}_{\mbox{\tiny II}}L^{(\mu)}_{\mbox{\tiny I}}
-L^{(e)}_AL^{(\mu)}_{\mbox{\tiny I}}
+L^{(e)}_{\mbox{\tiny II}}L^{(\mu)}_A\right)
\\
&
+\cos^2\theta(1+ 2\cos^2\theta)
\left(L^{(e)}_{\mbox{\tiny II}}L^{(\mu)}_{\mbox{\tiny I}}
-L^{(e)}_AL^{(\mu)}_{\mbox{\tiny II}}
-L^{(e)}_{\mbox{\tiny II}}L^{(\mu)}_A\right)\biggr]
+\frac{9}{16}\frac{mM}{M^2-m^2}\ln\frac{M^2}{m^2}\Biggr\}.
\end{split}
\eeq

\noindent
The first terms of the expansion of the contribution to HFS of the diagrams in Fig.~\ref{combo}  in the small mass ratio are already known for some time \cite{kp1951,egs2004}

\beq
\begin{split}
\Delta E
&=\frac{\alpha(Z^2\alpha)(Z\alpha)}{\pi^3}E_F
\biggl[
\frac{\pi^2}{2}\left(\ln{2}-\frac{13}{4}\right)
\\
&
+\frac{m}{M}\left(-\frac{9}{8}\ln{\frac{M}{m}} -\frac{15}{8}\zeta{(3)}+
\frac{15\pi^2}{4}\ln{2} +\frac{37\pi^2}{24}-
\frac{147}{32}\right)
+\frac{9}{16}
\frac{Mm}{M^2-m^2}\ln{\frac{M^2}{m^2}}
\biggr].
\end{split}
\eeq

\noindent
We have checked numerically that in the case of small mass ratio the expression in \eq{combogen} coincides with sum above.

In the case of equal masses the integral  in \eq{combogen} simplifies

\beq
\begin{split}
\Delta E_6
&
=\frac{\alpha^3}{\pi^3}E^{Ps}_F\Biggl\{ \frac{1}{4\pi} \int_0^\infty {dq^2\int_0^\pi d\theta\sin^2\theta d\theta}
\biggl[(2+\cos^2\theta )(L^2_{\mbox{\tiny I}}+2L_AL_{\mbox{\tiny I}})
\\
&
-6\cos^2\theta(L_{\mbox{\tiny I}}L_{\mbox{\tiny II}}+L_A L_{\mbox{\tiny II}}
-L_AL_{\mbox{\tiny I}})
+\cos^2\theta(1+ 2\cos^2\theta)
(L^2_{\mbox{\tiny II}}-2L_AL_{\mbox{\tiny II}})\biggr]
+\frac{9}{16}\Biggr\}.
\end{split}
\eeq

\noindent
After numerical calculations we obtain contribution of the diagrams in Fig.~\ref{combo} to HFS in positronium

\beq \label{combon}
\Delta E_6=-4.739~55~(40)~\frac{\alpha^3}{\pi^3} E^{Ps}_F.
\eeq

\section{Summary of Results}

We have derived explicit expressions for hard three-loop contributions to hyperfine splitting generated by the six gauge-invariant sets of diagrams with closed electron loops in Figs.~\ref{twooneloop} - \ref{onelopradph} and in Figs.~\ref{polinrad} - \ref{combo}. In the case of unequal lepton masses we confirm numerically the already known results for muonium obtained earlier in the form of an expansion in the small mass ratio.  We have calculated the contributions of these diagrams to HFS in the case of equal masses, for positronium. Collecting the results in \eq{onepol}, \eq{twpol}, \eq{polelf}, \eq{polinrd}, \eq{lbln}, and \eq{combon}, we  obtain the total hard contribution to HFS in positronium of order $m\alpha^7$ generated by all diagrams with closed electron loops in Figs.~\ref{twooneloop} - \ref{onelopradph} and in Figs.~\ref{polinrad} - \ref{combo}

\beq
\Delta E=-3.875~0~(4)\left(\frac{\alpha}{\pi}\right)^3 E^{Ps}_F=-1.291~7~(1)\frac{m\alpha^7}{\pi^3}=-5.672~\mbox{kHz}.
\eeq

Taking into account all other recent theoretical results \cite{bmp2014,af2014,apswf2014} we obtain the theoretical prediction for HFS in positronium

\beq
\Delta E_{theor}=203~391.89~(25)~\mbox{MHz}.
\eeq

\noindent
The latest experimental result is \cite{inaksyty2013}

\beq \label{expnew}
\Delta E_{exp}= 203~394.2~(1.6)_{stat}~(1.3)_{sys}~\mbox{MHz}.
\eeq

\noindent
There are no contradictions between theory and experiment at the present level of accuracy, but further reduction of both the experimental and theoretical uncertainties is warranted. Calculation of the remaining ultrasoft and hard nonlogarithmic contributions  of order $m\alpha^7$ is the next task for the theory. We hope to report the results for the remaining hard corrections in the near future.

\acknowledgments

This work was supported by the NSF grants PHY-1066054 and PHY-1402593. The work of V. S. was supported in part by the RFBR grant 14-02-00467.


\end{document}